# Small Q-D Neutrino Masses from a Generic Lepton Mass Hierarchy

E. M. Lipmanov

40 Wallingford Road #272, Brighton MA 02135, USA

Abstract

Exponential lepton mass ratios are studied in a low energy phenomenology. In view of the known data, the mass ratio patterns of the charged leptons (CL) and widely discussed quasi-degenerate (Q-D) neutrinos are related to one another by two different traits - opposite mass ratios ($x_n$) with large versus small exponents, and probably conformable mass-degeneracy-deviation ($x_n -1$) hierarchies. The solar-atmospheric hierarchy parameter ($\Delta m^2_{sol}/\Delta m^2_{atm}$) should have a special physical meaning in the Q-D scenario. A general generic hierarchy equation, with two opposite solutions for the CL and Q-D neutrino mass ratios, is considered. It determines a small upper bound on the neutrino mass scale with estimations $(m_\nu)_{max} \cong 0.30$ eV at 90% C.L., and $(m_\nu)_{max} \cong 0.18$ eV at best-fit mass-squared differences.

## 1. Introduction

The type of the neutrino mass spectrum – hierarchical, inverted or Q-D - is one of the major problems in neutrino physics. It is related to the problem of absolute neutrino masses and is beyond the neutrino oscillation experiments, which probe only the neutrino mass-squared differences. Direct measures of the absolute neutrino masses in tritium β-decay, neutrinoless double beta decay and by analyses of the cosmological data can solve the problem, but so far yielded only some upper limits on the values of the absolute neutrino masses.

By the available data, the noticeable phenomenological distinction of the Q-D neutrino mass pattern from the other ones come into view, in particular, by comparison with the CL mass pattern. The CL and Q-D neutrino mass spectra have opposite mass ratios – large and near to unity. Nevertheless, the mass degeneracy deviation hierarchies[1] of the two opposite lepton mass spectra may be conformable [1]. In contrast to the large disparity between the neutrino and CL mass scales and mass ratios, a conformity between their mass hierarchies is a remaining possible relation of the neutrino and CL mass patterns which is not in disagreement with the known to date lepton mass data.

In this Letter we consider inferences for absolute neutrino masses from a general form of neutrino-CL mass hierarchy analogy, extending [1].

In Sec.2, a possible conformity of mass hierarchies of the Q-D neutrinos and CL is considered. In Sec.3, a small upper bound on the Q-D neutrino mass scale is obtained in terms of oscillation mass-squared differences from a general quantitative neutrino-CL mass hierarchy analogy. The conclusions are given in Sec.4.

## 2. Conformable nonlinear mass hierarchies of the CL and Q-D neutrinos

By definition, the sequence of the neutrino masses $m_i$, i = 1,2,3, and mass ratios $x_n$, n = 1,2, let be

$$m_1 < m_2 < m_3, \quad x_1 \equiv m_2/m_1, \quad x_2 \equiv m_3/m_2. \tag{1}$$

Consider the known positive result of the neutrino oscillation experiments for the solar-atmospheric hierarchy parameter [2–4]:

$$(\Delta m^2_{sol}/\Delta m^2_{atm})_{exp} \equiv r \ll 1. \tag{2}$$

For Q-D neutrinos it follows[2]

$$\Delta m^2_{atm} \cong (x_2^2 - 1)m_\nu^2, \quad \Delta m^2_{sol} \cong (x_1^2 - 1)m_\nu^2, \tag{3}$$

where $m_\nu \cong m_2 \cong m_1$ is the neutrino mass scale, and

$$(\Delta m^2_{sol}/\Delta m^2_{atm}) \cong (x_1^2 - 1)/(x_2^2 - 1), \tag{4}$$

$$(x_1^2 - 1)/(x_2^2 - 1) \cong r \ll 1. \tag{5}$$

The dimensionless quantities $(x_n - 1)$, n = 1,2, estimate the relative deviations from the neutrino mass degeneracy. For an arbitrary Q-D neutrino mass pattern, and (5), it should be

$$x_1^2 = \exp \varepsilon_1, \quad x_2^2 = \exp \varepsilon_2, \quad \varepsilon_1 \cong r\, \varepsilon_2 \ll \varepsilon_2 \ll 1. \tag{6}$$

The small mass-ratio exponents $\varepsilon_1$ and $\varepsilon_2$ determine the relative (dimensionless) splitting of the neutrino masses.

---

[1] For short, the term "mass hierarchy" will be used at times in place of a more accurate phrase "nonlinear hierarchy of the dimensionless deviations from mass-degeneracy".

[2] An alternative connection of the solar and atmospheric mass-squared differences with the Q-D neutrino mass levels can also be chosen.

As a plain possibility, the main terms of the expansion series of the exponents $\varepsilon_2$ and $\varepsilon_1$ in powers of the small solar-atmospheric hierarchy parameter $r$ are given by

$$\varepsilon_2 \cong 2ar, \ \varepsilon_1 \cong 2ar^2. \qquad (7)$$

The unknown constant $a$ in (7) obeys the condition $0 < a \ll 1/r$, but should not necessarily be small or large. And so, the mass ratios of Q-D neutrinos can be represented in an exponential form

$$x_2 \cong \exp ar, \ x_1 \cong \exp ar^2. \qquad (8)$$

The nonlinear relation between the exponents in (8) is a possible result of both the near to unity Q-D neutrino mass ratios and the neutrino oscillation data (2).

The Q-D neutrino mass scale $m_\nu$ is given by

$$m_\nu^2 \cong \Delta m^2_{atm}/(2ar). \qquad (9)$$

The parameter $r$ has a special physical meaning in the Q-D neutrino scenario: 1). It is the definitely small phenomenological factor (2) in the mass-ratio exponents (8), 2). It is independent of the neutrino mass scale, 3). It measures the hierarchy of the deviations from neutrino mass degeneracy $r \cong (x_1 - 1)/(x_2 - 1) \ll 1$, 4). The condition $r \neq 0$ determines the (hierarchical) neutrino mass splitting, 5). By the known neutrino oscillation data, the value of the solar-atmospheric hierarchy parameter $r$ may be close to the value of the semi-weak analogue of the low energy fine structure constant [1], $r \cong \alpha_W = g_W^2/4\pi \cong 1/30$.

Like the parameter $r$, the coefficient $a$ in (8) is related to observable data,

$$a \cong \Delta m^2_{atm}/2(m_2^2)r, \qquad (10)$$

but in contrast to $r$, for an estimation of the coefficient $a$ one needs absolute neutrino mass data in addition to the oscillation data.

The WMAP cosmological upper bound on the Q-D neutrino mass scale [5], $m_\nu < 0.23$ eV (or a more conservative value [6], $m_\nu < 0.34$ eV), if taken at face value, leads to a lower bound on the coefficient $a$. For an estimation, with the best-fit atmospheric mass-squared difference [7],

$$\Delta m^2_{atm} = 2.0 \times 10^{-3} \text{ eV}^2, \qquad (11)$$

and the best-fit solar one [2,3,8],

$$\Delta m^2_{atm} = 7 \times 10^{-5} \text{ eV}^2, \qquad (12)$$

that lower bound on the coefficient $a$ is given by

$$a_{min} \cong 0.54 \ (m_\nu < 0.23 \text{ eV}); \ a_{min} \cong 0.25 \ (m_\nu < 0.34 \text{ eV}). \qquad (13)$$

The Q-D neutrino mass ratios (8) obey a nonlinear equation

$$(x_2^2 - 1)^2 \cong 2a(x_1^2 - 1), \qquad (14)$$

with a > $a_{min}$ from (13). It is a possible nonlinear hierarchy relation between the relative deviations from mass-degeneracy in a Q-D neutrino sector of the leptons.

Consider the known data for the CL masses [9]: $m_e \cong 0.511$ Mev, $m_\mu \cong 105.66$ Mev, $m_\tau \cong 1777$ Mev. Two large mass ratios $x_n$, n= 1,2, and a large mass-ratio hierarchy characterize the mass pattern of the CL

$$x_1 = m_\mu/m_e \gg 1, \quad x_2 = m_\tau/m_\mu \gg 1,$$
$$(m_\tau/m_\mu)^2 = \xi(m_\mu/m_e), \quad \xi \cong 1.37. \qquad (15)$$

Because of the large CL mass ratios, relation (15) can be rewritten in another form

$$[(m_\tau/m_\mu)^2 - 1]^2 \cong \xi^2[(m_\mu/m_e)^2 - 1]. \qquad (16)$$

This equation describes the nonlinear hierarchy of the very large relative deviations from mass-degeneracy in the CL sector of the leptons.

Compare the nonlinear Eqs. (14) and (16). They are conformable to one another: with $x_2 = m_\tau/m_\mu$, $x_1 = m_\mu/m_e$ and $\xi^2 = 2a$, Eq.(16) coincides with Eq.(14). The shift from the mass ratios ($x_n$) to the quantities ($x_n - 1$) makes possible the ansatz of conformable mass-degeneracy-deviation hierarchies of the CL and Q-D neutrinos.

The nonlinear equation (16) has an exponential solution,

$$x_1 = m_\mu/m_e \cong \xi \exp \chi, \quad x_2 = m_\tau/m_\mu \cong \xi \exp \chi/2, \quad \chi \gg 1, \qquad (17)$$

with one unknown parameter $\chi$. In this solution, the violation of the mass state (lepton flavor) symmetry is a large effect: $\chi_{exp} \cong \log[(m_\mu/m_e)/\xi] \cong 5.018$. The CL mass ratios are given by

$$m_\mu/m_e \cong \xi \exp 5, \quad m_\tau/m_\mu \cong \xi \exp 5/2, \qquad (18)$$

to within a percent. There is an interesting quantitative coincidence between the low energy semi-weak constant $\alpha_W$ and the integer 5: $\alpha_W \cong 5\exp(-5)$. If $r \cong \alpha_W$, the physical meaning of this coincidence is an approximate relation between the exponents $r$ and $\chi$, $r \cong \chi \exp(-\chi)$, as yet another possible connection between the lepton mass ratios (8) and (17). That relation is in agreement with the best-fit data values of the solar and atmospheric neutrino mass-squared differences (11) and (12) to within a few percent, and verifiable by the coming accurate neutrino oscillation data.

To conclude, the neutrino and CL mass ratios (8) and (17), with conformable mass-degeneracy-deviation hierarchies, are probable phenomenological characteristics of the complete lepton mass-ratio pattern if the Q-D neutrino scenario is indeed realized.

## 3. Small Q-D neutrino masses from a generic lepton mass hierarchy

Let us consider a generic lepton hierarchy equation as an extension of the two conformable hierarchy conditions (14) and (16) for the Q-D neutrinos and CL. The nonlinear CL hierarchy relation (15) does not change content when raised to an arbitrary power k. So, on account of the large CL mass ratios, a generic nonlinear lepton mass hierarchy equation in terms of mass-ratio powers k is given by

$$(x_2^k - 1)^2 \cong (\xi^k)(x_1^k - 1), \quad \xi_{exp} \cong 1.37, \quad k \geq 1. \quad (19)$$

At k = 2 it is reduced to Eq.(16), and to Eq.(14) with $a \cong \xi^2/2$. Equation (19) has two solutions with large and small exponents

$$m_\mu/m_e \cong \xi \exp \chi, \quad m_\tau/m_\mu \cong \xi \exp \chi/2, \quad \chi_{exp} \cong 5, \quad (20)$$

$$m_3/m_2 \cong \exp a_k r, \quad m_2/m_1 \cong \exp a_k r^2, \quad r_{exp} \cong (\Delta m^2_{sol}/\Delta m^2_{atm}), \quad (21)$$

for the CL and neutrino mass ratios (first power) respectively. These mass-ratio solutions are intertwined by the parameter $\xi$. The known very large violation of the CL mass-degeneracy and a possible weakly broken Q-D neutrino mass symmetry are mutually related by the generic Eq.(19). Such a relation between two solutions of the same equation will be termed *duality* relation in the present discussion. This duality relation can be substantiated by a possible connection between the small and large mass-ratio exponents $r$ and $\chi$, $r \cong \chi \exp(-\chi)$, Sec.2.

The approximate CL solution (20) does not depend on the power k in the Eq.(19), but the solution for the neutrino mass ratios does, the coefficient $a_k$ in (21) is a function of k,

$$a_k = \xi^k/k. \quad (22)$$

The neutrino solution (21) has the Q-D form (8), but the coefficient $a$ in this form is now a function of k, (22).

It should be noted that the functional dependence (22) of the coefficient $a_k$ on the mass-ratio power k is a consequence of the duality relation between the two exponential mass-ratio solutions of the hierarchy equation (19) for the CL and Q-D neutrinos.

The condition of a Q-D neutrino pattern,

$$a_k \ll 1/r_{exp}, \quad (23)$$

restricts also the allowed values of k ($1 < k \ll 1/r_{exp}$).

With the mass ratios (21), the Q-D neutrino mass scale is given by

$$m_\nu \cong (\Delta m^2_{atm}/2a_k r)^{1/2} \cong [\Delta m^2_{atm}/(x_2^2 - 1)]^{1/2}. \quad (24)$$

It is much more sensitive to the atmospheric neutrino mass-squared differences than to the solar ones.

The interesting point here is that the coefficient $a_k$ in (22), as a function of k has a unique, absolute minimum:

$$a_{min} = e \log\xi \cong 0.85 \qquad (25)$$

at $k = k_0 \equiv 1/\log\xi \cong 3.2$, where $e$ is the base of natural logarithms. This minimum is compatible with the estimations of the coefficient a in Sec.2. A statement is used: if $k_0$ minimizes the function $a_k = \xi^k/k$, we get $\xi^k|_{k=k0} = \exp 1 \equiv e$ independent of the value of the parameter $\xi > 1$. In accordance with that, the coefficient ($\xi^k$) in the equation (19) at the mass-ratio power $k = k_0$, i.e. at $a_k = a_{min}$ and $m_\nu = (m_\nu)_{max}$, is a known number $e = 2.718…$ and does not depend on the value of the coefficient $\xi$ in the initial relation (15).

The connection between the neutrino mass scale $m_\nu$ and the mass-ratio coefficient $a$, (9) or (24), transforms the minimum value $a_{min}$ (25) into the maximum value of the Q-D neutrino mass scale $(m_\nu)_{max}$,

$$(m_\nu)_{max} \cong \Delta m^2_{atm}/(2e \log\xi\, \Delta m^2_{sol})^{1/2}, \quad 2e \log\xi \cong 1.7, \qquad (26)$$
$$(m_\nu)_{max} \cong [\Delta m^2_{atm}/(x_2^2 - 1)_{min}]^{1/2},$$

since the neutrino mass-squared differences $\Delta m^2_{atm}$ and $\Delta m^2_{sol}$ are fixed and given by the neutrino oscillation data.

Note that the value (26) for $(m_\nu)_{max}$ is a little more than 20% larger than the value of the neutrino mass scale in the case $k = 1$; the maximum value $(m_\nu)_{max}$ of the neutrino mass scale is near to its value in case of mass-ratio squares $k = 2$.

If there is one generic equation for both the CL and Q-D neutrino mass hierarchies in accordance with (15), it should have the general form (19). If the value of the coefficient $a$ in (21), or (8), is changed (e.g. by some perturbation), but remains in the range

$$a_{min} \leq a \ll 1/r_{exp}, \qquad (27)$$

and the Q-D neutrino mass scale $m_\nu$ remains in the mass range

$$(\Delta m^2_{atm})^{1/2} < m_\nu \leq (m_\nu)_{max}, \qquad (28)$$

the neutrino mass ratios still obey equation (19) with a shifted value of the power k because the equation

$$a \cong \Delta m^2_{atm}/(2r\, m_\nu^2) = \xi^k/k \qquad (29)$$

has a solution for that power k. For an illustration, consider the case with $m_\nu = (m_\nu)_{max}$. The lepton mass ratios (20) and (21) with $k = k_0$ obey two different, though conformable equations $(x_2 - 1)^2 \cong \xi(x_1 - 1)$ and $(x_2 - 1)^2 \cong (e \log\xi)(x_1 - 1)$ for the CL and Q-D neutrinos respectively, but the $k_0$ powers of the mass-ratios (i.e. $x_n^{k0}$) of both the CL and neutrinos obey the same generic equation

$$(x_2^{k0} - 1)^2 \cong e(x_1^{k0} - 1), \quad k_0 = 1/\log\xi. \qquad (30)$$

For estimations, with the best-fit atmospheric mass-squared difference (11) and the best-fit one for the solar LMA MSW neutrino oscillation solution (12) we get from (24)

$$m_\nu \cong 0.14 \text{ eV}, \quad k = 1, \qquad (31)$$
$$m_\nu \cong 0.17 - 0.18 \text{ eV}, \quad 2 \leq k \leq 4, \qquad (32)$$

while the upper bound on the neutrino mass scale, from (26), is

$$(m_\nu)_{max} \cong 0.18 \text{ eV}. \qquad (33)$$

With the 90% C.L. ranges for the atmospheric and solar neutrino oscillations mass-squared differences [7,9],

$$\Delta m^2_{atm} \cong (1.3 - 3) \times 10^{-3} \text{ eV}^2, \qquad (34)$$
$$\Delta m^2_{sol} \cong (6 - 9) \times 10^{-5} \text{ eV}^2, \qquad (35)$$

we get from (24)

$$0.08 \text{ eV} < m_\nu \leq 0.24 \text{ eV}, \quad k = 1, \qquad (36)$$
$$0.1 \text{ eV} < m_\nu \leq 0.3 \text{ eV}, \quad 2 \leq k \leq 4, \qquad (37)$$

while with (26) the Q-D neutrino mass scale should be in the range

$$0.05 \text{ eV} < m_\nu \leq 0.3 \text{ eV}. \qquad (38)$$

These results for Q-D neutrinos are compatible with the constraints on neutrino masses from cosmological data analyses [5,6]. Eventually, further cosmological data could disprove the Q-D neutrino mass pattern if the upper constraint is less than $(\Delta m^2_{atm})^{1/2}$, or indeed prove that the neutrino mass scale is somewhere in the range (38).

If the condition of extreme neutrino mass-degeneracy-deviation values $(x_2 - 1)_{min} \cong 0.85r$ and $(x_1 - 1)_{min} \cong 0.85r^2$ determine the true Q-D neutrino mass scale $m_\nu$, rather than its maximum value, the neutrino mass scale is determined uniquely by the phenomenological equation (19) independent of the arbitrary value k of the mass-ratio powers involved and is given by

$$m_\nu \cong [\Delta m^2_{atm}/(x_2^2 - 1)_{min}]^{1/2} \cong [\Delta m^2_{atm}/1.7r]^{1/2}. \qquad (39)$$

The Q-D neutrino mass ratios are also uniquely inferred from equation (19):

$$m_3/m_2 \cong \exp(0.85r), \quad m_2/m_1 \cong \exp(0.85r^2). \qquad (40)$$

With the best-fit atmospheric and solar mass-squared differences (11) and (12), the estimation of the Q-D neutrino mass scale is

$$m_\nu \cong 0.18 \text{ eV}. \qquad (41)$$

With the 90% C.L. ranges of oscillation mass-squared differences (34) and (35), the Q-D neutrino mass scale is given by

$$0.11 \text{ eV} < m_\nu < 0.30 \text{ eV}. \qquad (42)$$

## 4. Conclusions

Exponential lepton mass ratios are studied in a low energy data oriented phenomenology. The Q-D neutrino type is singled out. The mass patterns of the Q-D neutrinos and CL may be related to one another by two different traits – opposite mass ratios $x_n$ and conformable mass-degeneracy-deviation $(x_n - 1)$ hierarchy equations. The special physical meaning of the solar-atmospheric hierarchy parameter $r = (\Delta m^2_{sol}/\Delta m^2_{atm})$ in the Q-D neutrino scenario is emphasized, a likely value $r \cong 1/30$ to within a few percent is noted. The ansatz of a generic hierarchy equation (19) determines an upper bound on the Q-D neutrino mass scale (26). The reason of its appearance and connection with the logarithm base $e$ is the duality relation of the two exponential solutions of the hierarchy equation (19) (*large versus small* exponents) for the CL and neutrinos, plus the CL data indication $\xi > 1$. With given small neutrino oscillation mass-squared differences, the value $(m_\nu)_{max}$ is determined by the condition of minimal hierarchical relative deviations from the neutrino mass degeneracy, $(x_1 - 1)_{min}$ and $(x_2 - 1)_{min}$. The estimations of the maximal value of the neutrino mass scale are $(m_\nu)_{max} \cong 0.30$ eV at 90% C.L. and $(m_\nu)_{max} \cong 0.18$ eV at the best-fit values of the neutrino oscillation mass-squared differences.